\documentclass[10pt]{revtex4}
\usepackage{graphicx}

\begin{document}

\title{Bose-Einstein condensates in `giant' toroidal magnetic traps}

 \author{A. S. Arnold and E. Riis}
 \affiliation{Department of Physics and Applied Physics, University of
Strathclyde, 107 Rottenrow, Glasgow G4 0NG, UK}
 \email{a.arnold@phys.strath.ac.uk}
 \homepage{http://www.photonics.phys.strath.ac.uk/People/Aidan/Aidan.html}
 \maketitle

 \noindent
\vspace{-8mm}

{\bf Abstract.} The experimental realisation of gaseous Bose-Einstein
condensation (BEC) in 1995 \cite{And1} sparked considerable interest in this
intriguing quantum fluid \cite{GSU}. Here we report on progress towards the
development of an $^{87}$Rb BEC experiment in a large ($\approx 10\,$cm
diameter) toroidal storage ring. A BEC will be formed at a localised region
within the toroidal magnetic trap, from whence it can be launched around the
torus. The benefits of the system are many-fold, as it should readily enable
detailed investigations of persistent currents, Josephson effects, phase
fluctuations and high-precision Sagnac or gravitational interferometry.

\vspace{3mm}

The idea \cite{ketttor} of neutral atomic and molecular storage rings has only
very recently been realised experimentally: fast polar molecules were stored in
a $25\,$cm diameter electrostatic hexapole ring \cite{natstor}, and
laser-cooled neutral $^{87}$Rb atoms have been confined in a $2\,$cm diameter
two-wire toroidal magnetic guide with an $\approx 1\,$s lifetime \cite{barr}.
We believe our `BEC-friendly' approach to toroidal neutral atom storage will
have additional advantages, ensuring that this new technology can also be used
to perform coherent atom optics with BECs. Our storage ring will be highly
adaptable, have a long $(\approx 1\,$minute) trapping lifetime, and have good
optical access (for both imaging and manipulation). In addition, as the BEC
will be prepared within the ring, coherently transferring the atoms into the
ring from a separate BEC creation site is not necessary.

\begin{figure}[!ht]
\begin{center}
\includegraphics[clip,width=15cm]{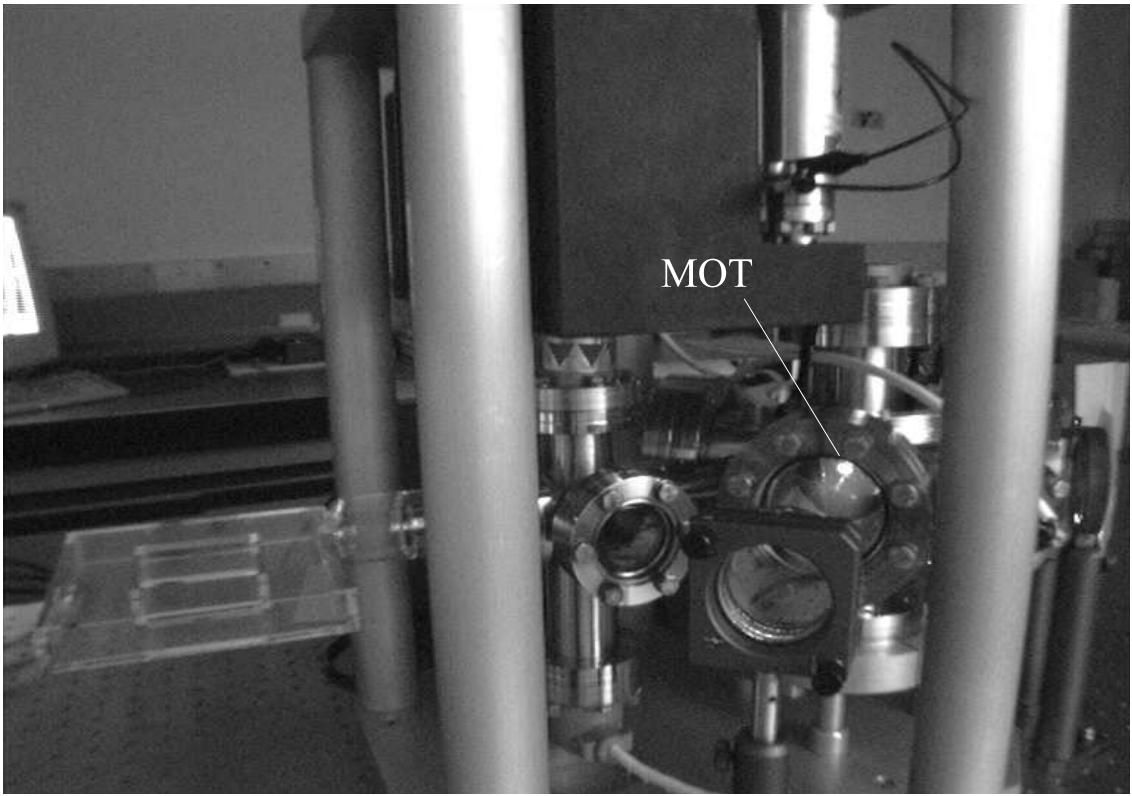}
\caption[]{\label{vac} The vacuum system consists of a `high pressure' conflat
cross (right) where an $N=2\times 10^{9}$ atom $^{87}$Rb MOT is now formed.
This MOT will be used to multiply load the special toroidal `low pressure'
quartz vacuum cell (enlarged in Fig.~\ref{cell}) which already yields pressures
less than $4\times 10^{-11}\,$torr.}
\end{center}
\end{figure}

A good atomic source can greatly lower the duration of BEC experiments. To this
end we have realised an $^{87}$Rb magneto-optical trap (MOT) \cite{raa} with a
relatively large atom number $(N=2\times10^{9}).$ The MOT was formed using low
powers $(40\,$mW) of trapping light generated by a simple diode laser system
\cite{arn}. The $780\,$nm $5s\,^{2}S_{1/2},F=2\rightarrow 5p\,^{2}P_{3/2},F=3$
trapping light is red-detuned by $20\,$MHz and comprised of three
retro-reflected two inch diameter laser beams. The loading rate of the MOT is
controlled by the current passing through SAES alkali metal dispensers
\cite{saes}. This MOT, situated in the high pressure end of the
differentially-pumped XHV vacuum chamber of Fig.~\ref{vac}, will be used as a
high-flux $(\approx 10^{10}$ atoms/s) source to multiply-load \cite{mya} a
second `low pressure' MOT in another chamber. We note that our atomic source
should yield comparable performance to a Zeeman slower \cite{zee}, without the
added complexity.

\begin{figure}[!ht]
\begin{center}
\includegraphics[clip,width=12cm]{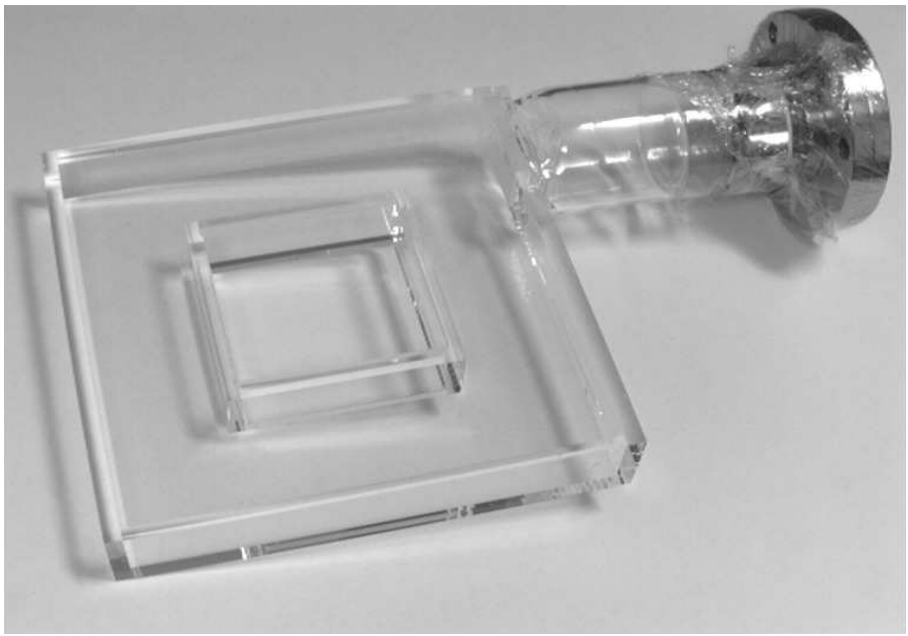}
\caption[]{\label{cell} The `square torus' high optical quality quartz vacuum
cell. The side-length of the square chamber is $12.5\,\mbox{cm}.$}
\end{center}
\end{figure}

\begin{figure}[!ht]
\begin{center}
\includegraphics[clip,width=15cm]{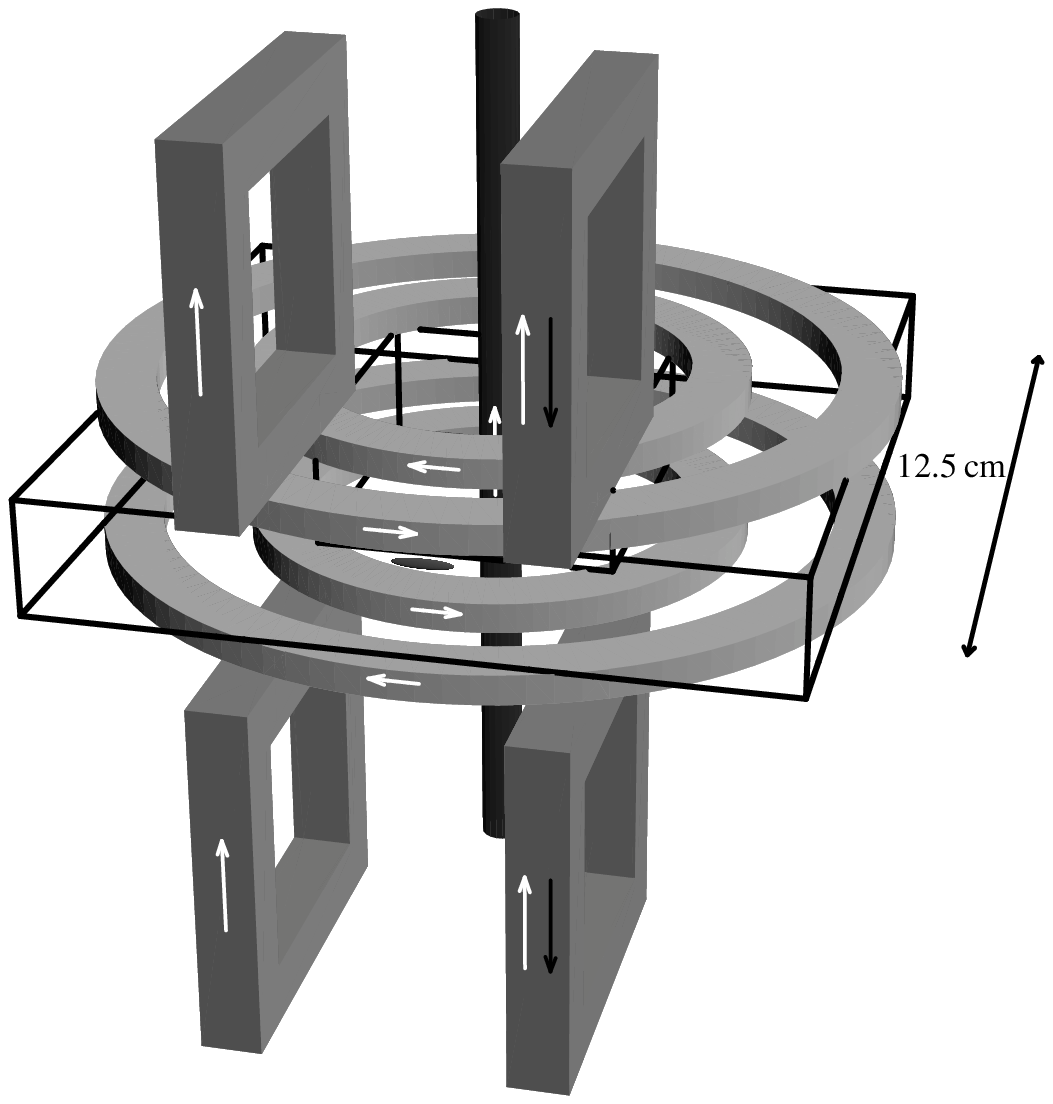}
\caption[]{\label{trap} The magnetic trap coils (to scale). Arrows indicate
current directions in the various coils. The small elliptical cloud represents
the axially elongated low pressure MOT/magnetic trap.}
\end{center}
\end{figure}

The `low pressure' MOT will be formed within a `square donut' quartz vacuum
cell (Fig.~\ref{cell}). This MOT will be an elongated \cite{dud}, forced dark
\cite{kett} MOT which should decrease light-assisted losses. The high product
of the MOT loading rate and lifetime will yield a large trapped atom
population. Ex-vacuo coils will be used for subsequent magnetic trapping of the
atoms in a localised region of the toroidal magnetic trap illustrated in
Fig.~\ref{trap}.

The currents of our magnetic coils (Fig.~\ref{trap}) will be regulated by five
water-cooled MOSFET banks \cite{mythesis}, and all currents will originate from
the same $5\,$V, $500\,$A power supply, ensuring optimal magnetic field noise
cancellation \cite{kettcurr} when the coils are connected in series
configurations. Our magnetic coil design allows us to realise many different
magnetic geometries. Quadrupolar toroidal confinement will be provided by the
circular coils (Fig.~\ref{trap}), which have mean diameters of $7.5$ and
$12.5\,$cm. Each coil carries a current of $2\,I_q=0-2\times 500\,$A, yielding
a maximum quadrupole magnetic field gradient of $227\,$G/cm. An azimuthal bias
field can be added at the minimum of the toroidal field using the black
straight wire, preventing spin-changing Majorana flops \cite{petr} in a purely
toroidal trap. The rectangular `pinch' coils in Fig.~\ref{trap} localise atoms
within the torus, and moreover they can be used in two distinct geometries,
namely quadrupole and Ioffe-Pritchard (IP) \cite{pri} configurations. The two
left rectangular coils are hard-wired in series, as are the right pair of
coils. These left and right pinch coil pairs carry currents of
$4\,I_{l}=0-4\times 500\,$A (with $I_l\leq I_q)$ and $4\,I_r=\pm 4 I_l$
respectively. The geometry of the rectangular coils is such that if $I_l=I_r$,
then in conjunction with the circular coils, an IP magnetic trap is formed. In
contrast, if $I_l=-I_r$ then the rectangular coils are equivalent to
anti-Helmholtz coils.

In the IP configuration the rectangular coils are designed to provide axial
curvature to the trap (up to $55\,$G/cm$^2$), whilst the axial bias field is
minimised ($\approx 0.9\,$G). Thus the optimal axial and radial magnetic
trapping frequencies are $10\,$Hz and $305\,$Hz respectively -- certainly
adequate for condensate formation \cite{usprl}. In the `anti-Helmholtz'
configuration the rectangular coils yield a maximum axial gradient of
$-67\,$G/cm (and the $227,-227\,$G/cm radial gradients from the toroidal
quadrupole coils change to $207,-140\,$G/cm).

As the same coils are used for the MOT and the IP magnetic trap, the centres of
the two magnetic fields overlap, and any possible `sloshing' effects during
transfers are eliminated. The adjustable nature of the axial magnetic field in
both MOT and IP trap configurations also enables excellent mode-matching
between the two traps.

The coils will initially be set to $I_q=I_l=-I_r=36\,$A, yielding radial and
axial magnetic field gradients of approximately $=15,-10\,$G/cm and $-5\,$G/cm
respectively. This provides an optimal field \cite{lind} for creating the `low
pressure' MOT. The aspect ratio of this MOT cloud will therefore be 0.7:1:1.2
(cf. 0.7:1:1 for a `standard' MOT). Once this MOT contains a few $10^9$ atoms,
then the MOT can be elongated axially by reducing the current $I_l=-I_r$ in the
rectangular coils. Atoms will not be collected directly in the elongated MOT
due to its shallow axial depth and short lifetime.

The atoms in the elongated MOT will then be released into optical molasses for
a few milliseconds, before being optically pumped into the
$|F,m_F\rangle=|2,2\rangle$ weak-field seeking state. Due to the elongated
geometry of the MOT, we expect that the atoms at this stage should have a
temperature as low as $10\,\mu$K, and a peak density of a few
$10^{11}\,$cm$^{-3}.$ By reversing the current $I_r,$ the atoms will then be
rapidly $(\ll 1\,$ms) loaded into the IP-type magnetic trap $(I_q=I_l=I_r)$
described above.

The magnetically trapped atoms can then be compressed by ramping up the current
in the magnetic coils. This increases the inter-atomic elastic collision rate
to facilitate efficient (`runaway') radio-frequency evaporation, which occurs
when the ratio of elastic:inelastic collisions is above $\approx$200:1
\cite{mythesis,luit}. The use of a `dark' MOT will probably not be necessary
for reaching the runaway evaporation threshold \cite{us}, however it will
increase the condensate population. If more magnetic compression is necessary,
current can be passed through the vertical wire (Fig.~\ref{trap}), and the MOT
can be loaded into an IP trap with an (initially) large bias field. After
implementing a suitable evaporation trajectory, condensation will occur in the
Ioffe-Pritchard trap in essentially the same manner as in most alkali BEC
experiments to date.

By ramping down the current in the magnetic trap's pinch coils, atoms can
subsequently access the entire torus, either by adiabatic expansion using the
pinch coils followed by non-adiabatic expansion, or using a Bragg launching
scheme. An important issue which needs to be addressed is the coherence of the
condensate. It can be shown \cite{us} that the coherence length of a
non-interacting isotropic BEC is given simply by the width of the condensate.
This result has also been seen experimentally along the `short' dimension of a
cigar-shaped interacting condensate \cite{kettref}. One might therefore
na\"{\i}vely expect that both the physical and coherence lengths of an axially
adiabatically expanding BEC would remain equal, especially as the `phase-space
density' of the condensate should remain well above one even if the condensate
is released non-adiabatically into the storage ring. However, recently
attention has been drawn to the issue of phase-fluctuations along the `long'
axis of cigar-shaped interacting BECs \cite{shly}, and such phase-fluctuations
have now been seen experimentally \cite{ert}. We also hope to study
phase-fluctuations, however their presence makes interferometry somewhat
difficult, and we will instead use a Bragg launching scheme (described below)
to initiate toroidal BEC motion.

Second-order Bragg scattering \cite{kettref,Phill} will be used to (multiply)
launch well-localised, {\it phase-coherent} atomic pulses (with very slowly
decaying phase-space density) out of the BEC, very shortly after the pinch
coils are turned off. One therefore avoids the large divergence of atomic
orbits that occur if one simply releases the BEC at the top of a vertically
oriented storage ring. The Bragg-scattered condensate pulses will be launched
in both directions around the ring. Due to their initial velocity of
$2.4\,$cm/s, one can only `lift' the BEC packets $h=30\,\mu$m vertically
against gravity (or up a magnetic field `hill' of $45\,$mG), and it is
therefore important to form the BEC at the `high' point of the toroidal
potential. For this reason we will tilt the axis of the toroidal trap with
respect to gravity, making the restrictions on magnetic field ripples less
stringent. Even vertical orientations of the magnetic ring are possible, as it
contains all atoms travelling at $<1.6\,$m/s.

Our `open' experimental geometry is well-suited to condensate investigation, in
particular for imaging the atoms. The high degree of optical access will also
facilitate the application of dipole force laser beams for manipulating the BEC
in persistent current/Josephson effect experiments. The Sagnac rotational
effect is proportional to the enclosed area of the beams in an interferometer.
The single-revolution enclosed area of our toroidal trap is $\approx140$cm$^2,$
which is more than six hundred times the area of the state-of-the-art thermal
atomic beam gyroscope \cite{gust}. Due to the long lifetime of our magnetic
trap, we envisage that BEC pulses could rotate around the storage ring hundreds
of times \cite{barr}, further enhancing the interferometer's sensitivity.
Additionally, it is hoped that the minimum detectable phase shift $(\delta \phi
\approx 1/\sqrt{N})$ decreases to $\delta \phi \approx 1/N$ \cite{jac} when
coherent atoms such as BECs are used, and higher precision Sagnac effect
measurement should therefore be possible. The magnetic trap will also provide a
good testing ground for investigations into the behaviour of one-dimensional
quantum gases.

\vspace{3mm}

{\bf Acknowledgments:} ASA is supported by a Royal Society of
Edinburgh/Scottish Executive Education and Lifelong Learning Department
research fellowship and the experiment is funded by the UK Engineering and
Physical Sciences Research Council.

\end{document}